\documentclass[aps,prl,twocolumn,floatfix,superscriptaddress,showpacs]{revtex4}

\usepackage{graphicx}
\usepackage{dcolumn}
\usepackage{bm}
\usepackage{textcomp}
\usepackage{color}

\begin{document}

\title{Spin splitting in graphene studied by means of tilted magnetic-field experiments}

\author{E.~V.~Kurganova}
\email[]{E.Kurganova@science.ru.nl}
\affiliation{
Radboud University Nijmegen, Institute for Molecules and Materials,High Field Magnet Laboratory,  Toernooiveld 7, 6525 ED Nijmegen, The Netherlands}
\author{H.~J. van Elferen}
\affiliation{
Radboud University Nijmegen, Institute for Molecules and Materials,High Field Magnet Laboratory,  Toernooiveld 7, 6525 ED Nijmegen, The Netherlands}
\author{A.~McCollam}
\affiliation{
Radboud University Nijmegen, Institute for Molecules and Materials,High Field Magnet Laboratory,  Toernooiveld 7, 6525 ED Nijmegen, The Netherlands}
\author{L.~A.~Ponomarenko}
\affiliation{
School of Physics and Astronomy, University of Manchester, M13 9PL, Manchester, United Kingdom
}
\author{K.~S.~Novoselov}
\affiliation{
School of Physics and Astronomy, University of Manchester, M13 9PL, Manchester, United Kingdom
}
\author{A.~Veligura}
\affiliation{
Physics of Nanodevices, Zernike Institute for Advanced Materials, University of Groningen, Nijenborgh 4, 9747 AG Groningen, The Netherlands
}

\author{B.~J. van Wees}
\affiliation{
Physics of Nanodevices, Zernike Institute for Advanced Materials, University of Groningen, Nijenborgh 4, 9747 AG Groningen, The Netherlands
}
\author{J.~C.~Maan}
\affiliation{
Radboud University Nijmegen, Institute for Molecules and Materials,High Field Magnet Laboratory,  Toernooiveld 7, 6525 ED Nijmegen, The Netherlands}
\author{U.~Zeitler}
\email[]{U.Zeitler@science.ru.nl}
\affiliation{
Radboud University Nijmegen, Institute for Molecules and Materials,High Field Magnet Laboratory,  Toernooiveld 7, 6525 ED Nijmegen, The Netherlands}

\date{\today}
\begin{abstract}
We have measured the spin splitting in single-layer and bilayer graphene by means of tilted magnetic field experiments.
Applying the Lifshitz-Kosevich formula for the spin-induced decrease of the Shubnikov de Haas amplitudes with increasing tilt angle we directly determine the product between the carrier cyclotron mass $m^{*}$ and the effective $g$-factor $g^{*}$ as a function of the charge carrier concentration. Using the cyclotron mass for a single-layer and a bilayer graphene we find an enhanced $g$-factor $g^{*}=2.7\pm0.2$ for both systems.
\end{abstract}

\maketitle

The half-integer quantum Hall effect in single-layer graphene (SLG) \cite{KostyaPioner,KimPioner} and the unconventional quantum Hall effect in bilayer graphene (BLG) \cite{BiUli} reveal spin- and valley-degenerate relativistic Landau levels.
Due to the extremely large Landau-level splitting~\cite{Giesbersgaps, Bilayergaps}, completely resolved levels can be observed up to room temperature \cite{RTQH}. However, even at very high perpendicular magnetic fields the Zeeman splitting within one Landau-level is negligible smaller compared to the Landau-level splitting and, more importantly, the Landau-level width generally exceeds the spin-splitting. Exceptionally, the zeroth Landau level in SLG becomes extremely narrow at magnetic fields $B>20$~T \cite{Giesbersgaps}, which allows an experimental observation of a spin-related gap opening at magnetic fields $B>20$~T \cite{Giesbersnu}.  Another observation of a spin degeneracy lifting with an effective $g$-factor $g^{*}=2$ was reported for $\nu=\pm4$, in SLG for magnetic fields $B>30$~T, combined with lifting the valley-degeneracy at $\nu=\pm1$ \cite{nu4Kim}.

In this paper we determine the spin splitting of broadened Landau levels for SLG and BLG by measuring Shubnikov-de Haas (SdH) oscillations in tilted magnetic fields. This technique allows adjusting the ratio between the spin splitting and the Landau level splitting, by controlling the ratio between a total magnetic field and a component perpendicular to a two-dimensional graphene flake. Using the well-established Lifshitz-Kosevich formula~\cite{LifshitzKosevich,AdamsHolstein} we determine the product of effective $g$-factor and cyclotron mass, $m^{*}g^{*}$,  from the angular dependence of the SdH amplitudes and we find that $g^{*}$ is enhanced compared to the free electron value.

We have fabricated field-effect transistors from SLG and BLG, by  micromechanically exfoliating graphene flakes from graphite. The flakes were deposited  on top of a Si/SiO$_{2}$ wafer, structured into a Hall-bar  and covered with Au/Ti contacts~\cite{Kostya2004}. Charge carriers are introduced by applying a gate voltage on the conducting Si substrate.

We present a detailed analysis on the spin splitting in a SLG sample made from Kish graphite with a mobility $\mu=0.8$~Vm$^{-2}$s$^{-1}$ and BLG sample originating from natural graphite with a mobility $\mu=0.3$~Vm$^{-2}$s$^{-1}$. Two other devices, one SLG and one BLG sample, showed qualitatively similar results.

\begin{figure}[tb]
  \begin{center}
  \includegraphics[width=0.9\linewidth]{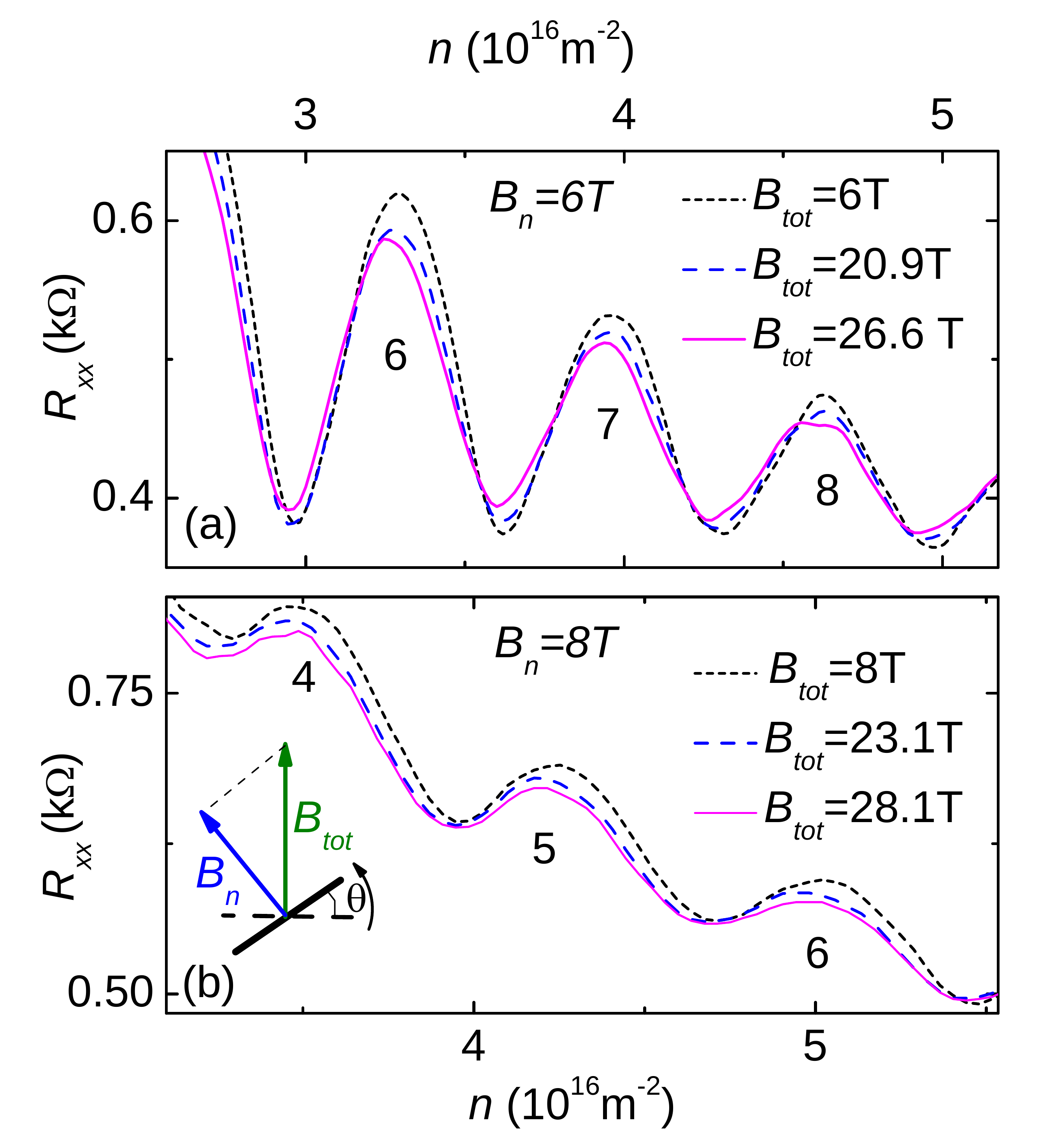}
  \end{center}
\vspace*{-1em}
\caption{(color online) Shubnikov de Haas oscillations in SLG (a) at $T=1.3$~K and in BLG(b) at $T=0.4~$K as a function of the carrier concentration for different total fields $B_{tot}$ or tilt angles $\theta$, respectively.
When varying $\theta$ the total field $B_{tot}$ is adjusted such that the perpendicular field $B_n$ remains constant, i.e.~ $B_{tot}= B_n / \cos \theta$. The oscillation maxima are marked with the corresponding Landau level numbers $N$. The inset schematically shows this tilting configuration.}
\label{TFFig1}
\end{figure}

To determine the spin-splitting we have measured the longitudinal resistances $R_{xx}$ as a function of charge carrier concentration $n$ at a constant perpendicular magnetic field. We adjusted the  total magnetic field $B_{tot}$ for each tilt angle such that the normal component $B_{n}$ is the same (inset to Fig.\ref{TFFig1}). The value of $B_{n}$ was verified by measuring the Hall resistance of the devices in the non-quantized regime.

In Fig.~\ref{TFFig1} we show the experimental $R_{xx}(n)$ dependencies for SLG at $B_{n}=6$~T (a) and for BLG at $B_{n}=8$~T (b). $R_{xx}$ shows Shubnikov-de Haas oscillations with maxima whenever the Fermi energy is situated in the middle of a spin- and valley-degenerated Landau level $E_N$, $N= 0,1,2, ...$ being the Landau-level index. For the higher Landau levels ($N\geq2$) the longitudinal resistances do not exhibit zero minima indicating that the level broadening is comparable to the cyclotron energy at these perpendicular magnetic fields.

When increasing $B_{tot}$ at a constant $B_n$ the oscillation amplitudes for both BLG and SLG are reduced. From this reduction we determined the spin-splitting. We use the Lifshitz-Kosevich formula  for systems with a general dispersion and we specifically include spin-splitting ~\cite{LifshitzKosevich,AdamsHolstein} with an effective $g$-factor $g^{*}$~\cite{Stephens,Dingle1} and tilted magnetic fields~\cite{Fang}. The oscillatory contribution to the longitudinal resistance can be described as \cite{KimPioner}:
\begin{equation}\label{LK}
\tilde{R}_{xx} =A~\cos \left( \frac{\hbar}{e B_{n}}\left.S(E)\right|_{E=E_F}+\pi+\varphi_{B} \right)
\end{equation}
where $\left.S(E)\right|_{E=E_F}$ is a extremal cross section of the Landau orbits in the $k$-space,  $A$ is the  oscillation amplitude and $\varphi_{B}$ is Berry phase, $\varphi_{B} = \pi$ for SLG \cite{KostyaPioner, KimPioner}, $\varphi_{B} = 2\pi$ for BLG \cite{BiUli}. The amplitude $A$ contains a monotonic $n$-dependent part, a temperature dependence, a $B_{n}$-dependent contribution and a damping factor due to spin splitting depending on the total field $B_{tot}$. At a constant temperature and perpendicular magnetic field this $B_{tot}$-dependence of the SdH amplitude $A$ for charge carriers with cyclotron mass $m^*$ and effective $g$-factor $g^*$ is given by \cite{Stephens,Fang}:
\begin{equation}\label{sf}
A=A_{0}(N) \cos \left( \frac{\pi}{2}\frac{g^{*}m^{*}}{m_{e}}\frac{B_{tot}}{B_{n}} \right)
\end{equation}
with cyclotron mass \cite{KostyaPioner}:
\begin{equation}\label{cycm}
\left.m^{*}=\frac{\hbar^{2}}{2\pi}\frac{dS(E)}{dE}\right|_{E=E_F}
\end{equation}
and $A_{0}(N)$ is constant for a given $N$.

For the spherical Fermi surface in SLG and BLG with a Fermi wave-vector $k_F = \sqrt{\pi n}$,  the extremal cross section of the Landau orbits is $\left.S(E)\right|_{E=E_F}=\pi k^{2}_{F} = n \pi^2$ and  Eq.~(\ref{LK}) yields the concentration-dependent resistance oscillations as we observe them in our experiments:
\begin{equation}
\label{LKn}
\tilde{R}_{xx}= A \cos \left( \frac{\hbar\pi^{2}}{eB_{n}}n+\pi+\varphi_{B} \right) = A \cos \left( \frac{\pi}{2} \nu +\pi+\varphi_{B} \right),
\end{equation}
where $\nu = (hn)/(eB_{n})$ is the filling factor. As expected, the oscillation period, $(2 e B_{n}) / (\hbar\pi)$,  is independent on the band structure of the 2D material and only depends on the filling factor.

\begin{figure}[tb]
  \begin{center}
  \includegraphics[width=0.9\linewidth,angle=0]{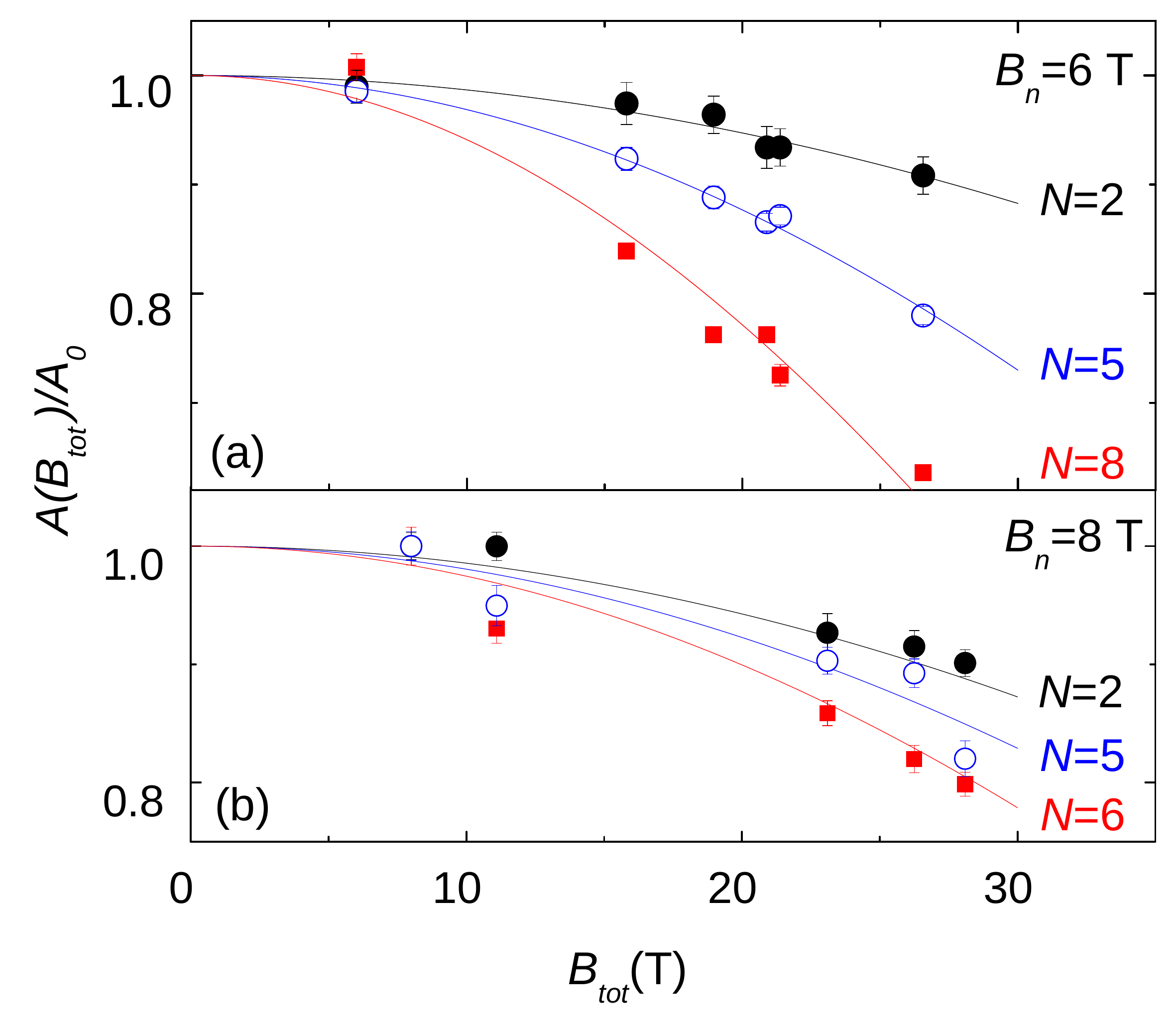}
  \end{center}
   \caption{(color online) Normalized oscillation amplitudes as a function of total field $B_{tot}$ at a constant perpendicular field $B_{n}$ in SLG (a) and BLG (b). Error bars represent standard least squares fitting errors in the determination of $A$. Solid lines are fits to Eq.~\ref{sf} with $m^*g^*$ as a fit parameter.}
\label{TFFig2}
\end{figure}

To accurately determine the experimental oscillation amplitudes  we have fitted our experimental data $R_{xx}(n)$ to Eq.~\ref{LKn} in two steps. First we determined the oscillation period and a smooth background using all oscillations measured for a wide range of the carrier concentrations. Second we fitted the oscillation amplitudes $A$ for each individual oscillation using the above determined period and background as  fixed parameters. In  Fig.~\ref{TFFig2} we show the final results of this fitting procedure for the  SdH amplitude as a function of the total magnetic field for different Landau levels $N$. For clarity all amplitudes are normalized to $A_{0}$.

\begin{figure}[tb]
  \begin{center}
  \includegraphics[width=0.9\linewidth]{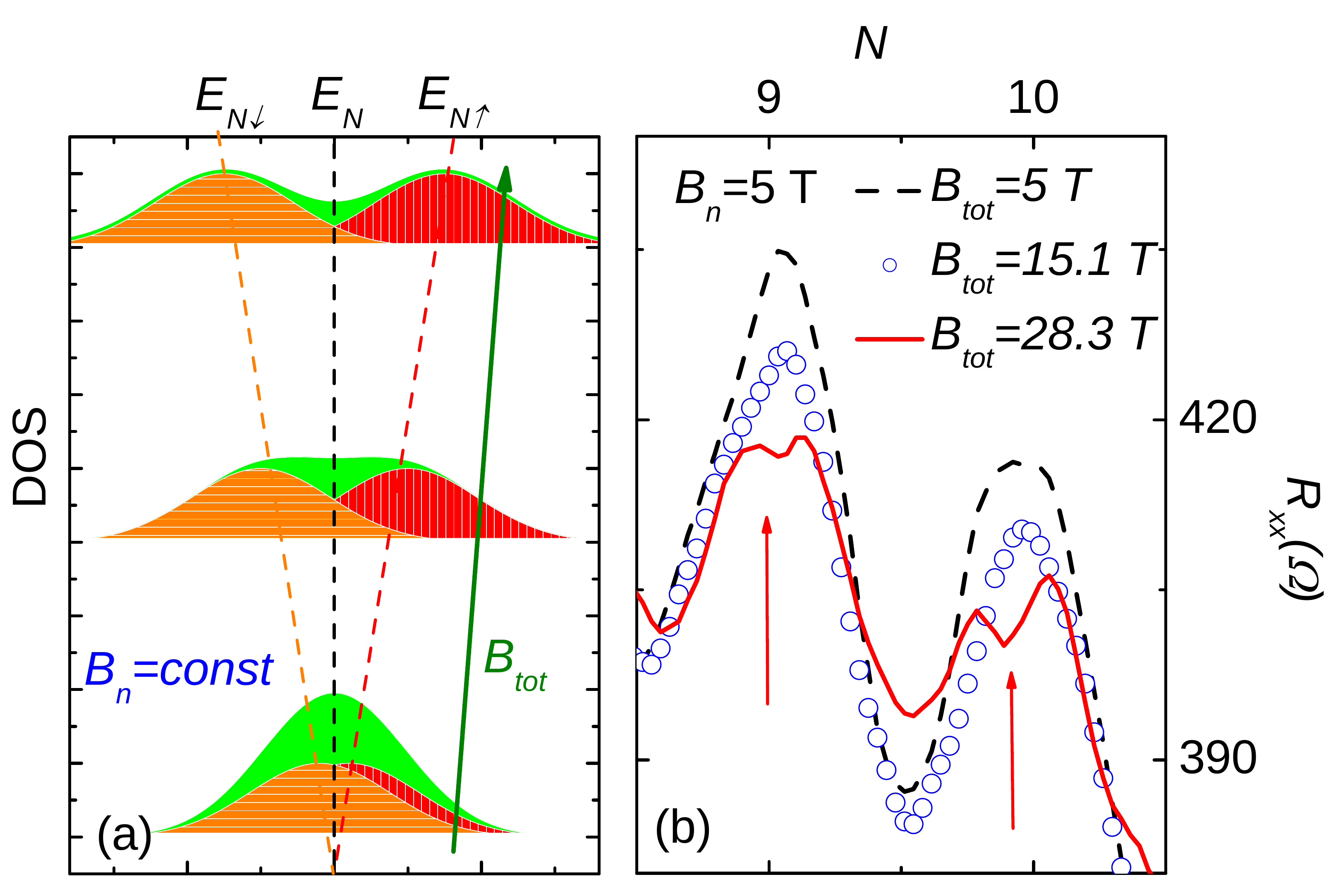}
  \end{center}
\vspace*{-1em}
\caption{(color online) Schematic representation of the density of states for a Landau level with an increasing total magnetic field $B_{tot}$ (from the bottom to the top) at a constant perpendicular component $B_{n}$ (a). Panel (b) shows this scenario as measured experimentally for the $N=9,10$ maximum in SLG at a constant perpendicular magnetic field $B_n = 5$~T.
}
\label{TFFig3}
\end{figure}

The experimentally observed reduction of the SdH amplitudes can be qualitatively visualized in a simple density of states (DOS) picture of a Landau level as depicted in Fig.~\ref{TFFig3}a. In a purely perpendicular magnetic field the Landau level width exceeds the spin splitting and the DOS of the spin-down state (orange, horizontally dashed in Fig.~\ref{TFFig3}a) overlaps with the one of the spin-up states (red, vertically dashed) to one broad Landau level. When increasing $B_{tot}$ by leaving $B_n$ constant, these two states move apart yielding an additional broadening of the Landau level with a reduced DOS in the center (green, solid areas in Fig.~\ref{TFFig3}a). Eventually, when the spin splitting exceeds the level width a minimum  between two distinct levels starts to develop in the DOS.
This scenario is indeed observed experimentally in SLG (Fig.~\ref{TFFig3}b). The SdH maxima corresponding to the $N=9$ and $N=10$ Landau levels at $B_{tot}=B_{n}$=5~T do not show any splitting. Increasing of the total field at a constant perpendicular component leads to a  reduction of the oscillation amplitude and  eventually appearance of spin-resolved peaks at the highest field of 28~T. However, this splitting is not yet enough to determine the energy difference by e.g. activation measurements.

\begin{figure}[tb]
  \begin{center}
  \includegraphics[width=0.9\linewidth,angle=0]{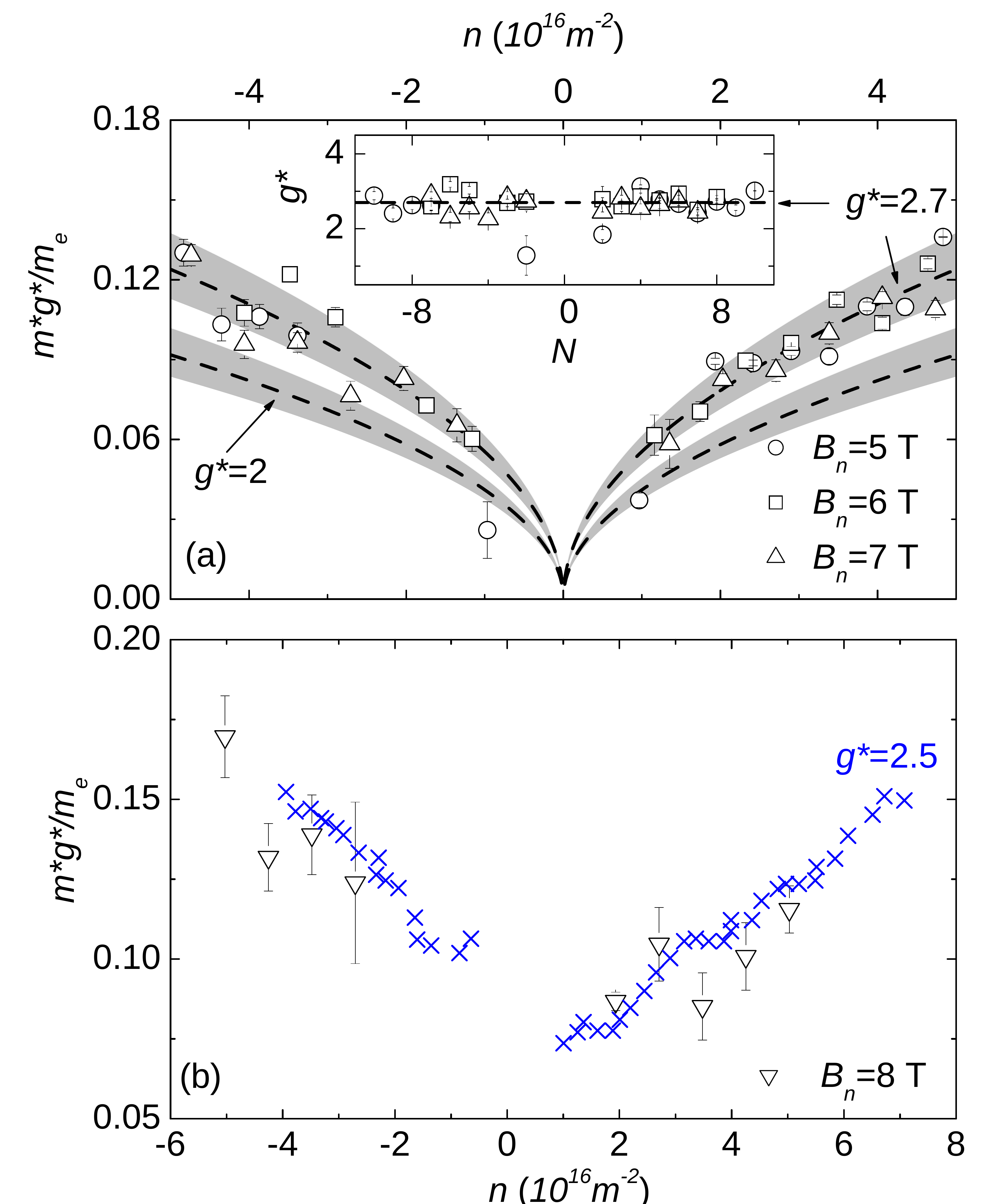}
  \end{center}
  \vspace*{-1em}
\caption{(color online) Experimentally deduced $m^{*}g^{*}$ (open symbols), normalized to the free electron mass $m_e$, as a function of charge carrier concentration for SLG (a) and BLG (b). The individual data points were extracted from the total-field dependence of the SdH amplitudes corresponding to different Landau levels $N= 2,...,10$ and represent measurement for a constant magnetic field $B_n = 5$~T, 6~T and 7~T for SLG and  $B_n = 8$~T for BLG. The error bars represent the standard least squares fitting errors, taking into account error bars of $A$ (Fig.~\ref{TFFig2}). The dashed lines in (a) represent the calculated behavior of $m^{*}g^{*}$ for different values of $g^{*}$  taking into account a 10\% experimental uncertainty (shadowed areas).
The  blue crosses in (b) compare our data to the experimental cyclotron mass for BLG~\cite{McycdopedBLG} multiplied by $g^{*}=2.5$.
The inset shows the effective g-factor, extracted from the product $m^{*}g^{*}$ in the main panel and the known cyclotron mass $m^*$ in SLG, as a function of Landau level index $N$.}
\label{TFFig4}
\end{figure}

A quantitative analysis of this decrease of the SdH amplitudes with increasing total magnetic field is done by fitting the data to Eq.~(\ref{sf}) with $m^{*}g^{*}$ as a fitting parameter (solid lines in Fig.~\ref{TFFig2}). The values for $m^{*}g^{*}$ obtained are plotted as a function of the charge carrier concentration in Fig.~\ref{TFFig4} for SLG (a) and BLG (b).

For both SLG and BLG the product $m^{*}g^{*}$ increases with concentration, which can be mainly attributed to the concentration dependent cyclotron mass $m^{*}$ of particles with a linear \cite{KostyaPioner} and hyperbolic dispersion \cite{Falko} as predicted by Eq.~\ref{cycm}.

The dashed lines in Fig.~\ref{TFFig4}a show the calculated dependence of $m^{*}g^{*}$ for $g^{*}=2$ and $g^{*}=2.7$ using $m^{*} (n) = (\hbar/c)\;\sqrt{\pi n}$ \cite{KostyaPioner}. The shadowed areas  represent a 10\% uncertainty of this calculation mainly due to the experimental errors and some uncertainty in the Fermi velocity~\cite{cfermi}.

For SLG (Fig.~\ref{TFFig4}a), the increase of $m^{*}g^{*}$ with $n$ is symmetric for electrons and holes (i.e.~negative and positive $n$ in the figure). A best fit using  $m^*(n)$ for SLG   yields $g^{*} = 2.7 \pm 0.2$ (the error is the standard deviation). This finding is shown directly in the inset of Fig.~\ref{TFFig4}a, where we plot the value of $g^{*}$ determined in the middle of each Landau level $N$ for different perpendicular fields $B_{n}$. Within an experimental error $g^{*}$ does not show any dependence on $N$ or $B_{n}$.

For BLG (Fig.~\ref{TFFig4}b) the experimental situation is more complex as the observed increase of $m^{*}g^{*}$ with $n$ is not symmetric for holes and electrons. Such a behavior is caused by an asymmetry of $m^*$ resulting from an asymmetric band structure of biased BLG, which was already observed experimentally in transport experiments~\cite{McycdopedBLG}, cyclotron resonance~\cite{Henriksen} and activation-gap measurements~\cite{Bilayergaps}. Applying the experimental cyclotron mass from Ref.~\onlinecite{McycdopedBLG} (depicted as blue crosses in Fig.\ref{TFFig4}) allows us to estimate $g^{*}$ to be about 2.5 for both electrons and holes which is, within experimental accuracy, reasonably consistent with the $g$-factor enhancement observed in SLG.

The observed enhancement of the effective spin-splitting compared to its free-electron value can be explained by electron-electron interaction~\cite{AndoUemura} yielding an interaction-enhanced splitting between two spin levels within one Landau level~\cite{Englert,Nicholas}:

\begin{equation}
\label{effSS}
g^{*}\mu_{B}B_{tot}=g\mu_{B}B_{tot}+E^{0}_{ex}(n_\downarrow-n_\uparrow).
\end{equation}
Here $g=2$ is a free-electron $g$-factor, $E^{0}_{ex}$ is an exchange parameter, and $n_\uparrow$ and $n_\downarrow$ are the relative occupations of the two spin states of a given Landau level.

For Gaussian shaped Landau levels with broadening \mbox {$\Gamma > g^{*}\mu_{B}B_{tot}$}, i.e.~where the spin splitting is not yet resolved, this relative occupation difference can be approximated using the Taylor expansion of the  Gauss error function ${\rm erf}(g^{*}\mu_{B}B_{tot}/\Gamma)$:
\begin{equation}
\label{spinpol}
n_\downarrow-n_\uparrow \approx \sqrt{\frac{1}{2\pi}} \; \frac{g^{*}\mu_{B}B_{tot}}{\Gamma}
\end{equation}
and Eq.(\ref{effSS}) yields:
\begin{equation}
\label{g-exc}
\frac{g^{*}}{g}=\left(1-\sqrt{\frac{1}{2\pi}}\frac{E^{0}_{ex}}{\Gamma}\right)^{-1}.
\end{equation}

$E^{0}_{ex}$ is of the order of Coulomb interaction, $E^{0}_{ex}\propto\sqrt{B_{n}}$~\cite{Nicholas}, and $\Gamma\propto \sqrt{B_{n}}$~\cite{Ando}. Therefore, the ratio $E^{0}_{ex}/ \Gamma$ remains constant and the $g$-factor enhancement is indeed predicted to be constant as we observe experimentally. Using the experimentally found $g^{*}=2.7$ in  Eq.~(\ref{g-exc}) yields $E^{0}_{ex}=130$~K at 10~T  when assuming $\Gamma=200$~K \cite{Giesbersgaps, Bilayergaps}.  For a completely spin polarized system, i.e. $n_\downarrow-n_\uparrow=1$, one might then speculate that the exchange enhancement in the Eq.~(\ref{effSS}) would be an order of magnitude larger than a single particle Zeeman energy at this particular field.

Finally, we note, that the experimentally found enhanced values of $g^{*}$ in graphene are close to those observed in transport experiments in graphite~\cite{Schneider}. This may suggest that an exchange induced enhancement of $g^{*}$ is quite common for graphitic materials. In contrast, no interaction-induced $g$-factor enhancement is observed using electron-spin resonance in graphene~\cite{ESRgraphene} and graphite~\cite{ESRgraphite} since these measurements are not sensitive to many body corrections~\cite{Kohn}.  Interestingly, measuring the Zeeman splitting of single-electron states in quantum dots, where no exchange enhancement of the $g$-factor is expected, also yields $g \approx 2$~\cite{Guttinger}, albeit with a considerable experimental uncertainty.

To conclude, we have experimentally measured and analyzed spin-splitting in SLG and BLG. We have shown that the product between the cyclotron mass $m^*$ and the effective $g$-factor $g^*$ increases with charge carrier concentration, as expected for a linear dispersion in SLG and a hyperbolic dispersion in BLG. Using the known concentration dependence of $m^*$  we found that $g^*$ in graphene is enhanced compared to the free-electron value and we attribute this to electron-electron interaction effects.

Part of this work has been supported by EuroMagNETII under EU contract 228043 and by the Stichting Fundamenteel Onderzoek der Materie (FOM) with financial support from the Nederlandse Organisatie voor Wetenschappelijk Onderzoek (NWO). The authors are grateful to Nacional de Grafite for supplying high quality crystals of graphite.



\end{document}